\documentclass[letterpaper,12pt]{article}
\usepackage[utf8]{inputenc}
\usepackage[margin=2cm]{geometry}
\usepackage{tcolorbox}
\usepackage{lipsum}
\usepackage{titling}
\usepackage[affil-it]{authblk} 
\usepackage[labelfont=bf]{caption}
\usepackage{textcomp}
\usepackage[colorlinks=true, allcolors=blue]{hyperref}
\usepackage{float}
\usepackage{amssymb}
\usepackage[arrowdel]{physics}
\usepackage{comment}
\usepackage{fancyhdr}
\usepackage{multicol}
\usepackage{makecell}
\usepackage{setspace}
\usepackage[normalem]{ulem}
\usepackage[style=chem-angew]{biblatex}
\title{{\it \textbf{\large{Effective Description of the Quantum Damped Harmonic Oscillator: Revisiting the Bateman Dual System}}}}

\author[1]{C. R. Javier Valdez\footnote{a310861@uach.mx}}
\author[1]{H. Hernandez-Hernandez\footnote{hhernandez@uach.mx}}
\author[2]{G. Chacon-Acosta\footnote{gchacon@cua.uam.mx}}

\affil[1]{Universidad Autonoma de Chihuahua, Facultad de Ingenieria, Nuevo Campus Universitario, Chihuahua 31125, Mexico}

\affil[2]{
Departmento de Matematicas Aplicadas y Sistemas, Universidad Autonoma Metropolitana-Cuajimalpa, Av. Vasco de Quiroga 4871, Ciudad de Mexico, 05348, Mexico}

\date{\today}

\begin{document}
\maketitle
\hrule

\begin{abstract}
In this work, we present a quantization scheme for the damped harmonic oscillator (QDHO) using a framework known as momentous quantum mechanics. Our method relies on a semiclassical dynamical system derived from an extended classical Hamiltonian, where the phase-space variables are given by expectation values of observables and quantum dispersions.
The significance of our study lies in its potential to serve as a foundational basis for the effective description of open quantum systems (OQS), and the description of dissipation in quantum mechanics. By employing the Bateman's dual model as the initial classical framework, and undergoing quantization, we demonstrate that our description aligns exceptionally well with the well-established Lindblad master equation. Furthermore, our approach exhibits robustness and broad applicability in the context of OQS, rendering it a versatile and powerful tool for studying various phenomena. We intend to contribute to the advancement of quantum physics by providing an effective means of quantizing the damped harmonic oscillator and shedding light on the behavior of open quantum systems.
 
\end{abstract}

\section{Introduction}

Quantum mechanics (QM) enables us to investigate the dynamics and interactions of physical systems
at atomic and subatomic scales. It has successfully addressed various phenomena
that classical physics could not explain. Notable examples include the black body
radiation, the double-slit, and the Stern-Gerlach experiments, among many others.
The traditional approaches of quantum mechanics, namely the Schrödinger
equation, or matrix mechanics, have primarily focused on closed systems, where no
 interaction with the environment occurs. However, quantum systems are not isolated but
constantly exchange energy and information with their surroundings: these are
known as open quantum systems (OQS). The significance of such systems has led
to the development of theoretical frameworks and computational techniques that
account for the complexities of real-world quantum phenomena.

The lack of a direct description of dissipation in quantum mechanics has been a
long-standing challenge, and usually simple systems are analyzed as they
serve as fundamental scenarios for investigation and could potentially offer insights
into more general methods. As such, the quantum damped harmonic oscillator (QDHO) has
been studied under several approaches \cite{DEKKER19811,um2002quantum,Um_1987}, that have, however, limitations to
varying extents. For example, for the Bateman dual system, the energy spectrum
fails to remain real-valued \cite{Feshbach_Tikochinsky,CHRUSCINSKI2006854}. Similarly, the time-dependent Caldirola-Kanai
Hamiltonian exhibits an exponential decay in the quantum analog, affecting the
evolution of the energy expectation value and position width \cite{Nassar2017,PhysRevA.55.935}, 
violating Heisenberg's uncertainty principle. These problems arise due to the
presence of non-Hermitian operators, responsible for dissipation, and their corresponding complex
eigenvalues, lacking a  physical interpretation \cite{Resonant_states}.

There are  alternative approaches to OQS using density matrix theory through
master equations, such as the Lindblad equation (also known as Gorini-Kossakowski-Lindblad-Sudarshan equation) 
 \cite{Lindblad,doi:10.1063/1.522979}, valuable schemes in studying OQS and their extensive applications in
quantum optics \cite{carmichael,walls_quantum_2008,scully_zubairy_1997,Quantum_Optics_Orzag}. However, master equations require certain assumptions
and approximations, such as the Born approximation, the Markov property, the
rotating wave approximation, weak coupling, and an infinite number of degrees of
freedom \cite{dekker_fundamental_1984}. This implies that, when utilizing master equations for analyzing more complex quantum systems, a thorough examination of these assumptions and a comprehensive understanding of their implications are essential. 

Frameworks like Bohmian mechanics and Everett's interpretation, presently without a clear treatment of open quantum systems, could benefit from incorporating a proper dissipation consideration. Both descriptions rely on the Schrödinger
equation (SE), leading again to the involvement of non-Hermitian operators \cite{SANZ20141,Vandyck_1994,Nassar2017}.

The utilization of effective theories for studying nontrivial quantum mechanical
systems has emerged as a compelling alternative \cite{Effective_Evolution}. Particularly interesting are
methods derived from geometric formulations, such as the momentous quantum
mechanics \cite{doi:10.1142/S0129055X06002772}, which offer valuable mathematical tools and intuitive insights for
interpreting complex phenomena. Noteworthy applications of this method include
the quantum pendulum, quantum tunneling, the double-slit experiment, and
various quantum cosmological scenarios \cite{doi:10.1063/1.4748550,10.1142,PhysRevA.98.063417,HernandezHernandez_2023,PhysRevD.84.043514,Quantum_Cosmology}. In this work, we analyze the
QDHO as a testing ground for applying
momentous quantum mechanics to OQS. By revisiting
Bateman's dual model, we derive a system of differential equations that yield the
semi-classical dynamics of expectation values of observables for the oscillator.
Comparing these equations with those arising from the Lindblad master equation,
we discuss the conditions under which the systems coincide. Our analysis
demonstrates that momentous quantum mechanics successfully overcomes the
limitations encountered with the quantization methods for the Bateman
Hamiltonian mentioned above. We intend to advance the
understanding of dissipation in open quantum systems and emphasize the potential of
momentous quantum mechanics as a powerful tool in quantum research. The
versatility of this approach opens exciting prospects for exploring a wide range of
quantum phenomena with enhanced accuracy and physical insight.

\section{Approaches for the Quantum Damped Harmonic Oscillator}

In this section, we review two approaches to the problem of dissipation for the QHO. It provides a starting point for our effective description. The Bateman Hamiltonian is historically one of the first attempts to study the quantization of the QDHO, and it is still under study \cite{celeghini_quantum_1992,schuch_connections_2017,deguchi_quantization_2020,blasone_batemans_2004}. The Lindblad master equation is a widely used approach in quantum optics \cite{carmichael} and in quantum information \cite{Nielsen,manzano_short_2020}.

\subsection{Bateman's Dual System}
\label{sec:Bateman's_Dual_System}
The classical equation for the linearly damped harmonic oscillator is given by
\begin{equation}
\label{eq:Usual_damped_HO}
    \ddot x+2\lambda\dot x+\omega_{o}^{2}x=0
\end{equation}
where $\lambda$ is the damping constant, and $\omega_{0}$ is the natural frequency of the oscillator. This equation of motion can be obtained from the two-dimensional Bateman Lagrangian
\begin{equation}
\label{eq:Lagrangian}
\begin{split}    
L(x,\dot x, y, \dot y)&=m\dot{x}\dot{y}+\lambda m (x\dot{y}-\dot{x}y)-kxy
\end{split}.
\end{equation}
An equation for the auxiliary variable $y(t)$ can also be obtained
\begin{equation}
    \ddot{y}-2\lambda\dot{y}+\omega_{0}^{2}y=0
\end{equation}
which represents a mirror image, or time reversed oscillator. This coupled $x-y$ system is conserved, because the energy dissipated by the $x$-oscillator is being absorbed by the $y$-oscillator \cite{razavy_classical_2005}.

For the quantum version we need the Hamiltonian. By using the Legendre transformation, we obtain the Bateman Hamiltonian
\begin{equation}
\label{eq:Bateman_Hamiltonian}
\begin{split}
H&=m\dot x\dot y +kxy=\frac{1}{m}p_{x}p_{y}+ \lambda( y p_{y}-xp_{x})+\Omega^{2} xy\\
\end{split}
\end{equation}
where $\Omega^{2}=\omega_{0}^{2}-\lambda^{2}$. 
The canonical momenta read
\begin{equation}
\label{eq:canonical_momenta}
    p_{y}=m(\dot x +\lambda x), \quad p_{x}=m(\dot y -\lambda y),
\end{equation}
and it can be seen that classical position and momenta are canonical
\begin{equation}
    \{x,p_{x}\}=\{y,p_{y}\}=1.
\end{equation}
The usual canonical momenta $p_{i}=m\dot{x}_{i},$ are modified due to dissipation, even in the limit $\lambda\to 0$.

To obtain the quantum dynamics,  promoting classical phase space variables to operators, we need to remove the ambiguity in the Hamiltonian (\ref{eq:Bateman_Hamiltonian}). To this end we use the Weyl ordering:
\begin{equation}
     W(\hat{x}_{i} \cdot \hat{p}_{i})=\frac{1}{2}(\hat{x}_{i}\hat{p}_{i}+\hat{p}_{i}\hat{x}_{i}).
\end{equation}

In this way we obtain the Hamiltonian operator for the Bateman model
\begin{equation}
\label{eq:op_Hamilton}
    \hat H = \frac{1}{m}\hat p_{x}\hat p_{y}-\frac{\lambda}{2}\Big((\hat x\hat p_{x}+\hat p_{x}\hat x)-(\hat y\hat p_{y}+\hat p_{y}\hat y)\Big)+\Omega^{2} \hat x\hat 
    y
\end{equation}
from which one obtains the quantum evolution of the QDHO. Following the quantization procedure of Feshbach and Tikochinsky \cite{Feshbach_Tikochinsky,razavy_classical_2005}, or the approach used by Chruścińsk and Jurkowski \cite{CHRUSCINSKI2006854}, one obtains the following energy spectrum
\begin{equation}
\label{eq:Batem_energy_eigen}
\begin{split}
    \hat H |\psi_{j,m}^{\pm}\rangle&=E_{j,m}^{\pm}|\psi_{j,m}^{\pm}\rangle\\
    &=\left(2\hbar\Omega j\pm i\hbar\lambda (2m+1)\right)|\psi_{j,m}^{\pm}\rangle, \quad m=|j|, |j|+1/2, |j|+1,...,\\
\end{split}
\end{equation}
$j \in \mathbb{Z}$. Thus, complex eigenvalues are obtained.
We can see that a unitary evolution, and a physical interpretation, is no longer possible. Moreover, as shown by Dekker in \cite{DEKKER19811}, the uncertainty relations decay to zero, violating Heisenberg's principle.
\subsection{The master equations approach: Lindblad Equation}
\label{sec:Master_Equations_Lindblad}
The Lindblad master equation is formulated within the Density Matrix Theory (DMT). This framework offers a set of tools allowing the study of pure and mixed states by using the density operator
\begin{equation}
    \rho:=\sum_{n} p_{n}|\psi_{n}\rangle\langle\psi_{n}|,
\end{equation}
where $|\psi_{n}\rangle$ is a normalized vector in the Hilbert space $\mathcal{H}$. It also applies to composite quantum systems by using the reduced matrix\footnote{For  a composite system given by $\rho_{AB}=\rho\otimes\sigma$, where $\rho$ and $\sigma$ belong to Hilbert spaces $\mathcal{H}_{A}$ and $\mathcal{H}_{B}$ respectively, the reduced density operator on the composite system $\rho_{AB}$ is
\begin{equation*}
\begin{split}
  \rho_{A}:&=\textrm{Tr}_{B}\{\rho_{AB}\}\\
  &=\textrm{Tr}_{B}\big\{\rho\otimes\sigma\big\}\\
  &=\rho \textrm{Tr}\{\sigma\}
\end{split}
\end{equation*}
} \cite{Nielsen,rivas_open_2012}.

 The basic idea to describe dissipation under DMT is by defining a total Hamiltonian $\hat{H}_{T}$
\begin{equation}
    \hat{H}_{T}=\hat{H}_{S}+\hat{H}_{R}+\hat{H}_{SR},
\end{equation}
composed by the system of interest $\hat{H}_{S}$, the environment or reservoir $\hat{H}_{R}$, and the interaction between them $\hat{H}_{SR}$, forming a closed system that can be analyzed by the von Neumann equation. Because, in general, $\hat{H}_{T}$ describes an extremely complex system, the problem is put in a more tractable form by using the reduced matrix method, which allows the study of a subsystem of the composite system, thus limiting the analysis only to the system of interest $\hat{H}_{S}$.

$\hat{H}_{S}$, $\hat{H}_{R}$ and the interaction $\hat{H}_{SR}$ are defined as follows
\begin{align}
        &\hat{H}_{S}=\hbar\omega \hat{a}^{\dagger}\hat{a},\nonumber\\
        &\hat{H}_{R}=\sum_{j}\hbar\omega_{j}\hat{r}_{j}^{\dagger}\hat{r}_{j},\nonumber\\
        &H_{SR}=\sum_{j}\hbar(k_{j}^{*}\hat{a}\hat{r}_{j}^{\dagger}+k_{j}\hat{a}^{\dagger}\hat{r}_{j}).
\end{align}
The system of interest $\hat{H}_{S}$ is the QHO, with $\hat{a}^{\dagger}$ and $\hat{a}$ are the creation and annihilation operators, respectively. The reservoir Hamiltonian $\hat{H}_{R}$ is modeled by an infinite number of harmonic oscillators, where $\hat{r}_{j}^{\dagger}$ and $\hat{r}_{j}$ are the corresponding creation and annihilation operators of the $jth$ oscillator. Finally, $\hat{H}_{SR}$ is the interaction between the system of interest and the reservoir, and $k_{j}$ are coupling constants.

Following \cite{carmichael,walls_quantum_2008,Quantum_Optics_Orzag}, where the Born, Markov, rotating wave, and weak coupling approximations are used, one arrives at the Lindblad master equation for the QDHO
\begin{equation}
\label{Quantum Damped Harmonic Oscillator}
    \dot{\rho}=-i\omega_{o}'[a^{\dagger}a,\rho]+\frac{\gamma}{2}(2a\rho a^{\dagger}-a^{\dagger}a\rho-\rho a^{\dagger}a)+\gamma\bar{n}(a\rho a^{\dagger}+a^{\dagger}\rho a-a^{\dagger}a\rho-\rho aa^{\dagger}),
\end{equation}
where $\omega_{o}'$ is a frequency shift $\omega_{o}+\Delta$, $\gamma$ is a damping constant, and $\bar{n}=\bar{n}(\omega,T)$ is the mean photon number for the reservoir oscillators in thermal equilibrium at temperature $T$.

Instead of solving this equation for $\rho(t)$, as was done in \cite{fujii2013quantum}, one can work directly with the evolution of expectation values of observables as in \cite{SANDULESCU1987277}
\begin{equation}
\label{eq:expectation_val_eq}
    \frac{d}{dt}\langle \hat{O} \rangle=Tr\{\hat{O}\dot{\rho}\}.
\end{equation}
From it the mean energy evolution is obtained
 \begin{equation}
 \label{eq:mean_energy_lindblad}
     \langle \hat{E}(t) \rangle= \left\langle \hat{n}(t)+\frac{1}{2} \right\rangle\hbar\omega=\left(\big(\langle\hat{n}(0)\rangle-\bar{n}\big)e^{-\gamma t}+\bar{n}+\frac{1}{2}\right)\hbar\omega.
 \end{equation}
As $t \to \infty$  the energy of the system decays to an excited state above the ground state $(\bar{n}+1/2)\hbar\omega$. Focusing on the evolution of the expectation value of position and momentum operators
\begin{equation}
    \hat{x}=\sqrt{\frac{\hbar}{2m\omega}}\big(\hat{a}+\hat{a}^{\dagger}\big),\quad \hat{p}=\frac{1}{i}\sqrt{\frac{\hbar m\omega}{2}}\big(\hat{a}-\hat{a}^{\dagger}\big),
\end{equation}
equations of motion follow
\begin{equation}
\label{eq:class_lindblad_din}
\begin{split}
    \frac{d}{dt}\langle \hat{x} \rangle=\frac{\omega_{o}'}{m\omega}\langle \hat{p} \rangle-\frac{\gamma}{2}\langle \hat{x} \rangle,\quad \frac{d}{dt}\langle \hat{p} \rangle=-m\omega\omega_{o} '\langle \hat{x} \rangle-\frac{\gamma}{2}\langle \hat{p} \rangle,
\end{split}
\end{equation}
which are the classical equations of the damped oscillator. Thus, in the classical limit, the Lindblad master equation recovers the classical dynamics. To complement the quantum dynamics, we need to investigate the evolution of the dispersions $\langle \hat{x}^{2}\rangle$, $\langle \hat{p}^{2}\rangle$, which are obtained from Eqs. (\ref{eq:expectation_val_eq}) and (\ref{Quantum Damped Harmonic Oscillator})  
\begin{align}
\label{eq:system_lindblad_quantum}
    &\frac{d}{dt}\langle \hat{x}^{2} \rangle=-\gamma\langle \hat{x}^{2} \rangle+\frac{\omega_{o}'}{m\omega}\langle \hat{x}\hat{p}+\hat{p}\hat{x} \rangle+\frac{\gamma\hbar}{2m\omega}(2\bar{n}+1),\nonumber\\
    &\frac{d}{dt}\langle \hat{p}^{2} \rangle=-\gamma\langle \hat{p}^{2} \rangle- m\omega_{o}'\omega\langle \hat{x}\hat{p}+\hat{p}\hat{x} \rangle+\frac{\gamma\hbar m\omega}{2}(2\bar{n}+1),\nonumber\\
    &\frac{d}{dt}\langle \hat{x}\hat{p}+\hat{p}\hat{x} \rangle=-2m\omega_{o}'\omega\langle \hat{x}^{2} \rangle+\frac{2\omega_{o}'}{m\omega}\langle \hat{p}^{2} \rangle-\gamma\langle \hat{x}\hat{p}+\hat{p}\hat{x} \rangle,
\end{align}
From Eqs. (\ref{eq:class_lindblad_din}) and (\ref{eq:system_lindblad_quantum}) we observe that the classical $(\langle \hat{x} \rangle, \langle \hat{p} \rangle)$ and quantum $(\langle \hat{x}^2 \rangle, \langle \hat{p}^2 \rangle, \langle \hat{x} \hat{p} \rangle, \langle \hat{p} \hat{x} \rangle)$ variables decouple, thus, the classical evolution does not get modified by the quantum back-reaction \cite{doi:10.1142/S0129055X06002772,Quantum_Cosmology}.

\section{Effective description} \label{section:effective}

\subsection{Momentous Quantum Mechanics}
\label{section:formalism_momentous}

Momentous quantum mechanics \cite{doi:10.1142/S0129055X06002772,Quantum_Cosmology}, is a semiclassical setting that allows studying complex quantum systems by approximating their behavior through effective classical equations. Classical and quantum evolution are related by means of the following prescription
\begin{equation}
\label{eq: quantum_phase_space}
\{\langle\hat{A}\rangle,\langle\hat{B}\rangle\}=\frac{1}{i\hbar}\langle[\hat{A},\hat{B}]\rangle.
\end{equation}
Although expectation values of observables give classical variables, $x=\langle\hat{x}\rangle, p=\langle\hat{p}\rangle$, because in general $\langle\hat{A}^{n}\rangle\neq\langle\hat{A}\rangle^{n}$, most expectation values of quantum operators cannot be associated with classical variables. In the momentous formalism, the expectation value of general quantum correlation operators are defined as
\begin{equation}
\label{Generalized effective dynamical variables}
\begin{split}
    G^{a_{1},b_{1},\dots,a_{k},b_{k}}:=&\big\langle (\hat x_{1} -\langle\hat{x}_{1}\rangle)^{a_{1}}(\hat {p}_{1} -\langle\hat{p}_{1}\rangle)^{b_{1}}\cdots (\hat {x}_{k} -\langle\hat{x}_{k}\rangle)^{a_{k}}(\hat{p}_{k} -\langle\hat{p}_{k}\rangle)^{b_{k}}\big\rangle_{\textrm{Weyl}},
\end{split}
\end{equation}
for a system with $k$ degrees of freedom. Once more, Weyl ordering is employed. From this definition, for one degree of freedom, we can recover the usual fluctuations $(\Delta x)^{2}=G^{2,0}, (\Delta p)^{2}=G^{0,2}$, and covariance $ \Delta (xp)=G^{1,1}$. Heisenberg's uncertainty can be written in terms of these variables
\begin{equation}
\label{eq:uncertainty}
    G^{2,0}G^{0,2}-(G^{1,1})^{2}\geq\frac{\hbar^2}{4}.
\end{equation}

Equipped with this structure, and classical and quantum variables, the evolution of the quantum system can be obtained from the effective Hamiltonian $H_{Q}$
\begin{align}
\label{eq:H_q}
    \langle\hat{H}\rangle=H_{Q}&=\langle H[x_{1}+(\hat{x}_{1}-x_{1}),p_{1}+(\hat{p}_{1}-p_{1}),...,x_{k}+(\hat{x}_{k}-x_{k}),p_{k}+(\hat{p}_{k}-p_{k})]\rangle_{\text{Weyl}}\nonumber\\
    &=\sum_{a_{1},b_{1},...,a_{k},b_{k}}^{\infty}\frac{1}{a_{1}!b_{1}!\cdots a_{k}!b_{k}!}\frac{\partial^{ a_{1}+b_{1}+\cdots+a_{k}+b_{k}}H_{class}}{\partial x_{1}^{a_{1}}\partial p_{1}^{b_{1}}\cdots\partial x_{k}^{a_{k}}\partial p_{k}^{b_{k}}}G^{a_{1},b_{1},\dots,a_{k},b_{k}}\nonumber\\
    &=H_{class}+\sum_{a_{1}+b_{1}+\cdots+a_{k}+b_{k}\geq 2}^{\infty}\frac{1}{a_{1}!b_{1}!\cdots a_{k}!b_{k}!}\frac{\partial^{ a_{1}+b_{1}+\cdots+a_{k}+b_{k}}H_{class}}{\partial x_{1}^{a_{1}}\partial p_{1}^{b_{1}}\cdots\partial x_{k}^{a_{k}}\partial p_{k}^{b_{k}}}G^{a_{1},b_{1},\dots,a_{k},b_{k}}
\end{align}
with $H_{class}=H(x_{1},p_{1},...,x_{k},p_{k})$. Equations of motion can be obtained by obtaining the Poisson bracket between variables and the Hamiltonian\footnote{For instance, the algebra of quantum variables up to second order, for one degree of freedom, is \[
    \{G^{2,0},G^{1,1}\}=2G^{2,0},\quad \{G^{2,0},G^{0,2}\}=4G^{1,1}, \quad \{G^{1,1},G^{0,2}\}=2G^{0,2} 
\]}
\begin{equation}
\label{eq: Hamilton_formalism}
    \frac{d\langle \hat{f}\rangle}{dt}=\{\langle \hat{f}\rangle,H_{Q}\},
\end{equation}
considering that classical and quantum variables are simplectic orthogonal
\begin{equation}
\label{eq:difference_quantum_classical}
    \{x_{k},G^{a_{1},b_{1},...,a_{k},b_{k}}\}=\{p_{k},G^{a_{1},b_{1},...,a_{k},b_{k}}\}=0.
\end{equation}

The above Hamiltonian can be understood as a classical one augmented with quantum corrections, and the resulting equations of motion provide an effective evolution equivalent to the Schrödinger equation \cite{ashtekar_geometrical_1997,doi:10.1142/S0129055X06002772}.
 In other words, the semiclassical description shows how quantum corrections modify the classical dynamics. We can always obtain the classical dynamics 
 in the classical limit $\hbar \rightarrow 0, \ G^{a_{1},b_{1},...,a_{k},b_{k}} \rightarrow 0$. 

\subsection{Effective Bateman Model}
\label{sec:semiclassical_equation_BT}
We now study the quantum evolution of the Bateman model shown in section \ref{sec:Bateman's_Dual_System}. Specifically, we use the Tikochinsky transformation for the Bateman Hamiltonian \cite{doi:10.1063/1.523718}, and then analyze its similarities with the total Hamiltonian used in the Lindblad master equation approach.

The Bateman-Tikochinsky Hamiltonian (BTH) is
\begin{equation}
\label{eq:bateman-tiko}
\begin{split}
H&=\left(\frac{p_{1}^{2}}{2m}+\frac{1}{2}m\Omega^{2}x_{1}^{2}\right)-\left(\frac{p_{2}^{2}}{2m}+\frac{1}{2}m\Omega^{2}x_{2}^{2}\right)-\lambda(x_{1}p_{2}+x_{2}p_{1}).\\
\end{split}
\end{equation}
Note the parallelism between both models as shown in section \ref{sec:Master_Equations_Lindblad}
\begin{center}
\text{Bateman-Tikochinsky}\quad\quad\quad\quad \text{Lindblad $H_{T}$}
\end{center}
\begin{align}
   &H_{S} & &\left(\frac{p_{1}^{2}}{2m}+\frac{1}{2}m\Omega^{2}x_{1}^{2}\right) & &\hbar\omega a^{\dagger}a\\
   &H_{R} & &\left(\frac{p_{2}^{2}}{2m}+\frac{1}{2}m\Omega^{2}x_{2}^{2}\right) & &\sum_{j}\hbar\omega_{j}r_{j}^{\dagger}r_{j}\\
   &H_{SR} & &\lambda(x_{1}p_{2}+x_{2}p_{1}) & &\sum_{j}\hbar(k_{j}^{*}ar_{j}^{\dagger}+k_{j}a^{\dagger}r_{j})
\end{align}
with the obvious differences regarding the number of degrees of freedom.

We can obtain the BTH by applying the following canonical transformations on Eq. (\ref{eq:Bateman_Hamiltonian}) for position and momenta
\begin{align}
\label{eq:trans}
    &x=\frac{1}{\sqrt{2}}\big(x_{1}+x_{2}\big), & &y=\frac{1}{\sqrt{2}}\big(x_{1}-x_{2}\big), \nonumber \\
    &p_{y}=\frac{1}{\sqrt{2}}\big(p_{1}-p_{2}\big), &  &p_{x}=\frac{1}{\sqrt{2}}\big(p_{1}+p_{2}\big).
\end{align}
As we showed in section \ref{sec:Bateman's_Dual_System}, the  corresponding canonical quantization generates an inconsistent physical evolution. 

We have found, however, a canonical transformation for the classical variables Eq. (\ref{eq:trans})
\begin{align} \label{eq:class_bateman2}
        &x_{1}\longrightarrow\hat{x}_{1}=\frac{\sqrt{2}}{2}\big(\hat{x}+\hat{y}\big),\quad p_{1}\longrightarrow\hat{p}_{1}=\frac{\sqrt{2}}{2}\big(\hat{p}_{x}+\hat{p}_{y}\big), \nonumber \\
         &x_{2}\longrightarrow\hat{x}_{2}=\frac{\sqrt{2}}{2}\big(\hat{x}-\hat{y}\big),\quad p_{2}\longrightarrow\hat{p}_{2}=-\frac{\sqrt{2}}{2}\big(\hat{p}_{x}-\hat{p}_{y}\big),        
\end{align}
whose quantum operators obey canonical commutation relations
\begin{equation} \label{bateman2}
    [\hat{x}_{1},\hat{p}_{1}]=[\hat{p}_{2},\hat{x}_{2}]=i\hbar,\quad\text{and}\quad [\hat{x}_{1},\hat{p}_{2}]=[\hat{x}_{2},\hat{p}_{1}]=0. 
\end{equation}
that indeed provides a physically correct evolution.
The quantum dynamical variables for the BTH are given by
\begin{equation}
    G_{1}^{a,b,c,d}:=\Big\langle\big(\hat{x}_{1}-\langle\hat{x}_{1}\rangle\big)^{a}\big(\hat{p}_{1}-\langle\hat{p}_{1}\rangle\big)^{b}\big(\hat{p}_{2}-\langle\hat{p}_{2}\rangle\big)^{c}\big(\hat{x}_{2}-\langle\hat{x}_{2}\rangle\big)^{d}\Big\rangle_{\text{Weyl}}
\end{equation}
and the corresponding quantum variables are as follows
\begin{align*} 
    G^{2,0,0,0}&=\Big\langle \big(\hat{x}-\langle \hat{x}\rangle\big)^{2} \Big\rangle_{\text{Weyl}} & G^{0,2,0,0}&=\Big\langle \big(\hat{p}_{x}-\langle \hat{p}_{x}\rangle\big)^{2} \Big\rangle_{\text{Weyl}}\\
    &=\frac{1}{2}\Big\langle \big[(\hat{x}_{1}-\langle \hat{x}_{1}\rangle)+(\hat{x}_{2}-\langle \hat{x}_{2}\rangle)\big]^{2} \Big\rangle_{\text{Weyl}} & &=\frac{1}{2}\Big\langle \big[(\hat{p}_{1}-\langle \hat{p}_{1}\rangle)-(\hat{p}_{2}-\langle \hat{p}_{2}\rangle)\big]^{2} \Big\rangle_{\text{Weyl}}\\
    &=\frac{1}{2}\Big[G^{2,0,0,0}_{1}+G^{0,0,0,2}_{1}+2G^{1,0,0,1}_{1}\Big] & &=\frac{1}{2}\Big[G^{0,2,0,0}_{1}+G^{0,0,2,0}_{1}-2G^{0,1,1,0}_{1}\Big]
\end{align*}
\begin{align} \label{eq:quantum_bateman2}
    G^{1,1,0,0}&=\Big\langle \big(\hat{x}-\langle x\rangle\big)\big(\hat{p}_{x}-\langle p_{x}\rangle\big) \Big\rangle_{\text{Weyl}}\nonumber\\
        &=\frac{1}{2}\Big\langle \big[(\hat{x}_{1}-\langle x_{1}\rangle)+(\hat{x}_{2}-\langle x_{2}\rangle)\big]\big[(\hat{p}_{1}-\langle p_{1}\rangle)-(\hat{p}_{2}-\langle p_{2}\rangle)\big] \Big\rangle_{\text{Weyl}}\nonumber\\
        &=\frac{1}{2}\Big[G^{1,1,0,0}_{1}-G^{0,0,1,1}_{1}-G^{1,0,1,0}_{1}+G^{0,1,0,1}_{1}\Big]
\end{align}

Classical variables (\ref{eq:trans}), quantum variables (\ref{eq:class_bateman2}), and the BTH Eq.(\ref{eq:bateman-tiko}), give the quantum corrected Hamiltonian $\langle H_{Q} \rangle$, Eq. (\ref{eq:H_q}) 
\begin{align}
\label{eq:quantum_corrected}
    \langle H_{Q} \rangle&=H_{classical}+H_{quantum}\nonumber\\
    &=\left(\frac{p_{1}^{2}}{2m}+\frac{1}{2}m\Omega^{2}x_{1}^{2}\right)-\left(\frac{p_{2}^{2}}{2m}+\frac{1}{2}m\Omega^{2}x_{2}^{2}\right)-\lambda(x_{1}p_{2}+x_{2}p_{1})-\lambda G^{1,0,1,0}_{1}\nonumber\\
    &+\frac{m\Omega^{2}}{2}G^{2,0,0,0}_{1}+\frac{1}{2m}G^{0,2,0,0}_{1}-\frac{m\Omega^{2}}{2}G^{0,0,0,2}_{1}-\frac{1}{2m}G^{0,0,2,0}_{1}-\lambda G^{0,1,0,1}_{1}.
\end{align}

Henceforth we will call this Hamiltonian, Eq. (\ref{eq:quantum_corrected}), the Semiclassical Bateman-Tikochinsky Hamiltonian (SBTH).

We are ready to study the effective evolution. Using Eqs. (\ref{eq: quantum_phase_space}) and (\ref{eq: Hamilton_formalism}) we obtain the dynamical equations of motion
\begin{align*}
    &\dot{x}_{1}=\frac{p_{1}}{m}-\lambda x_{2},\nonumber\\ &\dot{x}_{2}=-\frac{p_{2}}{m}-\lambda x_{1},\nonumber\\
        &\dot{p_{1}}=-m\Omega^{2}x_{1}+\lambda p_{2},\nonumber\\ &\dot{p_{2}}=m\Omega^{2}x_{2}+\lambda p_{1},\nonumber\\
        &\dot{G}^{2,0,0,0}_{1}=-2\lambda G_{1}^{1,0,0,1}+\frac{2}{m}G_{1}^{1,1,0,0},\nonumber\\
        &\dot{G}^{0,2,0,0}_{1}=2\lambda G_{1}^{0,1,1,0}-2m\Omega^{2}G_{1}^{1,1,0,0},\nonumber\\
        &\dot{G}^{0,0,2,0}_{1}=-2\lambda G_{1}^{0,1,1,0}-2m\Omega^{2}G_{1}^{0,0,1,1},\nonumber\\
        &\dot{G}^{0,0,0,2}_{1}=2\lambda G_{1}^{1,0,0,1}+\frac{2}{m}G_{1}^{0,0,1,1},\nonumber\\
        &\dot{G}^{1,0,1,0}_{1}=-\lambda G_{1}^{1,1,0,0}-\lambda G_{1}^{0,0,1,1}+\frac{1}{m}G_{1}^{0,1,1,0}-m\Omega^{2}G_{1}^{1,0,0,1},\nonumber\\
        &\dot{G}^{0,1,0,1}_{1}=\lambda G_{1}^{1,1,0,0}+\lambda G_{1}^{0,0,1,1}-\frac{1}{m}G_{1}^{0,1,1,0}-m\Omega^{2}G_{1}^{1,0,0,1},\nonumber\\
        &\dot{G}^{1,0,0,1}_{1}=\lambda G_{1}^{2,0,0,0}-\lambda G_{1}^{0,0,0,2}+\frac{1}{m}G_{1}^{0,1,0,1}+\frac{1}{m}G_{1}^{1,0,1,0},\nonumber\\
\end{align*}
\begin{align}
\label{eq:QDHO_sys_eq} 
        &\dot{G}^{0,1,1,0}_{1}=\lambda G_{1}^{0,0,2,0}-\lambda G_{1}^{0,2,0,0}-m\Omega^{2}G_{1}^{1,0,1,0}-m\Omega^{2}G_{1}^{0,1,0,1},\nonumber\\
        &\dot{G}^{1,1,0,0}_{1}=\lambda G_{1}^{1,0,1,0}-\lambda G_{1}^{0,1,0,1}+\frac{1}{m}G_{1}^{0,2,0,0}-m\Omega^{2}G_{1}^{2,0,0,0},\nonumber\\
        &\dot{G}^{0,0,1,1}_{1}=\lambda G_{1}^{1,0,1,0}-\lambda G_{1}^{0,1,0,1}+\frac{1}{m}G_{1}^{0,0,2,0}-m\Omega^{2}G_{1}^{0,0,0,2},
\end{align}

If we were to restore to original variables for the Bateman model, Eq. (\ref{eq:Usual_damped_HO}), we could rewrite Eq. (\ref{eq:QDHO_sys_eq}), resulting in a very interesting SDE
\begin{align} 
\label{eq:G_QDHO}
    &\dot{x}=\frac{1}{m}p_{x}-\lambda x\nonumber\\
    &\dot{p}_{x}=-m\Omega^{2}x-\lambda p_{x}\nonumber\\
    &\dot{G}^{2,0,0,0}=-2\lambda G^{2,0,0,0}+\frac{2}{m}G^{1,1,0,0}+\frac{2}{m}\big[G_{1}^{0,0,1,1}+G_{1}^{1,0,1,0}\big]+2\lambda\big[G_{1}^{2,0,0,0}+G_{1}^{1,0,0,1}\big]\nonumber\\
    &\dot{G}^{0,2,0,0}=-2\lambda G^{0,2,0,0}-2m\Omega^{2}G^{1,1,0,0}+2\lambda\big[G_{1}^{0,2,0,0}-G_{1}^{0,1,1,0}\big]-2m\Omega^{2}\big[G_{1}^{0,0,1,1}-G_{1}^{0,1,0,1}\big]\nonumber\\
    &\dot{G}^{1,1,0,0}=-2\lambda G^{1,1,0,0}+\frac{1}{m}G^{0,2,0,0}-m\Omega^{2}G^{2,0,0,0}+\lambda[2G_{1}^{1,1,0,0}+G_{1}^{0,1,0,1}-G_{1}^{1,0,1,0}]\nonumber\\
    &\quad\quad\quad-\frac{1}{m}G_{1}^{0,0,2,0}+m\Omega^{2}[G_{1}^{0,0,0,2}+G_{1}^{1,0,0,1}],
\end{align}
where, the classical dynamics is decoupled from the quantum dynamics, and the latter is very similar to the one given in Eq. (\ref{eq:system_lindblad_quantum}). Let us stress exactly this by 
comparing the classical effective dynamics for the Lindblad and the momentous description
\begin{align} \label{eq:classic_equality1}
    &\quad\quad\text{Lindblad}\quad\quad\quad\quad\quad\quad\quad\quad\quad\quad\quad\text{SBTH}\nonumber\\
    &\frac{d}{dt}\langle \hat{x} \rangle=\frac{\omega_{o}'}{m\omega}\langle \hat{p} \rangle-\frac{\gamma}{2}\langle \hat{x} \rangle, \quad\quad\quad\quad \dot{x}=\frac{1}{m}p_{x}-\lambda x\nonumber\\
    &\frac{d}{dt}\langle \hat{p} \rangle=-m\omega\omega_{o} '\langle \hat{x} \rangle-\frac{\gamma}{2}\langle \hat{p} \rangle, \quad\quad\text{ } \dot{p}_{x}=-m\Omega^{2}x-\lambda p_{x}.
\end{align}
Both descriptions are equivalent if we set $\omega_{0}'=\omega=\Omega$ and $\gamma=2\lambda$. For the quantum counterpart, it is useful to rewrite the SDE for Lindblad in the following way 
\begin{align} \label{eq:quantum_equality}
    \frac{d}{dt}G_{L}^{2,0}&=\frac{d}{dt}\langle\hat{x}^{2}\rangle-\frac{d}{dt}\langle\hat{x}\rangle^{2}\nonumber\\
    &=-\gamma\langle \hat{x}^{2} \rangle+\frac{\omega_{o}'}{m\omega}\langle \hat{x}\hat{p}+\hat{p}\hat{x} \rangle+\frac{\gamma\hbar}{2m\omega}(2\bar{n}+1)-2\langle\hat{x}\rangle\frac{d}{dt}\langle\hat{x}\rangle\nonumber\\
    &=-\gamma\big(\langle \hat{x}^{2} \rangle-\langle\hat{x}\rangle^{2}\big)+2\frac{\omega_{o}'}{m\omega}\Big(\frac{\langle \hat{x}\hat{p}+\hat{p}\hat{x} \rangle}{2}-\langle\hat{x}\rangle\langle \hat{p} \rangle\Big)+\frac{\gamma\hbar}{2m\omega}(2\bar{n}+1)\nonumber\\
   \dot{G}_{L}^{2,0} &=-\gamma G_{L}^{2,0}+\frac{2\omega_{o}'}{m\omega}G_{L}^{1,1}+\frac{\gamma\hbar}{2m\omega}(2\bar{n}+1),
\end{align}
where $G_{L}$ is seen as a quantum variable in the momentous scheme. The full set of equations gives the Lindblad SDE, Eq. (\ref{eq:system_lindblad_quantum}), in terms of quantum variables in the semiclassical approach
\begin{align}
\label{eq:G_Lindblad}
        &\dot{G}_{L}^{2,0}=-\gamma G_{L}^{2,0}+\frac{2\omega_{o}'}{m\omega}G_{L}^{1,1}+\frac{\gamma\hbar}{2m\omega}(2\bar{n}+1)\nonumber\\
        &\dot{G}_{L}^{0,2}=-\gamma G_{L}^{0,2}-2m\omega\omega_{o}'G_{L}^{1,1}+\frac{\gamma\hbar m\omega}{2}(2\bar{n}+1)\nonumber\\
        &\dot{G}_{L}^{1,1}=-m\omega\omega_{o}'G_{L}^{2,0}+\frac{\omega_{o}'}{m\omega}G_{L}^{0,2}-\gamma G_{L}^{1,1}.
\end{align}

If we call the constant terms in the above equations in the following way %
\begin{equation}
D_{xx}=\frac{\gamma\hbar}{2m\omega}(2\bar{n}+1)\quad D_{pp}=\frac{\gamma\hbar m\omega}{2}(2\bar{n}+1)\quad D_{px}=0
\end{equation}
we can see that they obey the fundamental constrain for diffusion coefficients \cite{dekker_fundamental_1984,dekker_master_1979,isar_density_1993}
\begin{equation}
    D_{xx}>0\quad D_{pp}>0\quad D_{xx}D_{pp}-D_{px}^{2}\geq\frac{\hbar^{2}\gamma^{2}}{4}.
\end{equation}
Now, if in Eq. (\ref{eq:G_QDHO}) we perform a similar identification
\begin{equation}
    D_{Gxx}=2\lambda G_{1}^{2,0,0,0}\quad D_{Gpp}=2\lambda G_{1}^{0,2,0,0}\quad D_{Gpx}=2\lambda G_{1}^{1,1,0,0},
\end{equation}
 and taking into account the generalized uncertainty relation for quantum variables given in Eq. (\ref{eq:uncertainty})
\begin{equation}
    D_{Gxx}D_{Gpp}-D_{Gpx}^{2}\geq\lambda^{2}\hbar^{2},
\end{equation}
we get a complete agreement in both descriptions: for the SDE, Eq. (\ref{eq:G_QDHO}) also stands as the fundamental constraint for diffusion coefficients.

\section{QDHO effective evolution}

\subsection{Initial Conditions}

In order to study the evolution of the QDHO, Eq, (\ref{eq:G_QDHO}), we need to establish the initial conditions. We assume that, initially, no correlation exist between the physical oscillator and the reservoir. We also propose a coherent state as our initial wave function, for which initial conditions can be computed. 

The initial coherent state reads
\begin{equation}
    |\bar{x}_{0}\rangle=e^{-\frac{i}{\hbar}\hat{p}x_{0}}|0\rangle,
\end{equation}
yielding the following initial conditions for the quantum variables%
\begin{align}
\label{eq:initial_con_quantum_var}
    &(\Delta \hat{x})^{2}_{t=0}=\frac{\hbar}{2m\omega},  &(\Delta \hat{p})^{2}_{t=0}=\frac{m\hbar\omega}{2}, \nonumber\\   
    &\langle\hat{x}^{2}\rangle_{t=0}=\frac{\hbar}{2m\omega}+x_{0}^{2},   &\langle\hat{p}^{2}\rangle_{t=0}=\frac{m\hbar\omega}{2}, \nonumber\\
    &\langle\hat{x}\hat{p}+\hat{p}\hat{x}\rangle_{t=0}=0,    &\langle\hat{x}\rangle_{t=0}=x_{0}=\sqrt{\frac{2n\hbar}{m\omega}}.
\end{align}
 where $\langle \bar{x}_{0}|\hat{H}|\bar{x}_{0}\rangle=\frac{1}{2m}\langle \bar{x}_{0}|\hat{p}^{2}|\bar{x}_{0}\rangle+\frac{1}{2}m\omega^{2}\langle\bar{x}_{0}|\hat{x}^{2}|\bar{x}_{0}\rangle.$
Since the left hand side $\langle \bar{x}_{0}|\hat{H}|\bar{x}_{0}\rangle$ is one of the quantized energy levels of the QHO, say $\Big(n+\frac{1}{2}\Big)\hbar\omega$, we have
\begin{align}
    \Big(n+\frac{1}{2}\Big)\hbar\omega&=\frac{1}{2}\hbar\omega+\frac{1}{2}m\omega^{2}x^{2}_{0}\quad\longrightarrow\quad x_{0}=\sqrt{\frac{2n\hbar}{m\omega}}
\end{align}

 The corresponding conditions for the transformed variables in Eq. (\ref{eq:QDHO_sys_eq}) read
\begin{align}
            &x_{0}=\sqrt{\frac{2n\hbar}{m\omega}} & &y_{0}=\sqrt{\frac{2n\hbar}{m\omega}} & &\longrightarrow  & &\langle\hat{x}_{1}\rangle_{t=0}=x_{10}=2\sqrt{\frac{n\hbar}{m\omega}} & &\langle\hat{x}_{2}\rangle_{t=0}=x_{20}=0, \nonumber \\
            &\langle\hat{p}_{x}\rangle_{t=0}=p_{x_{0}}=0 & &\langle\hat{p}_{y}\rangle_{t=0}=p_{y_{0}}=0 & &\longrightarrow & &\langle\hat{p}_{1}\rangle_{t=0}=p_{10}=0 & &\langle\hat{p}_{2}\rangle_{t=0}=p_{20}=0.
\end{align}
The evolution for the mirror system is obtained with the same initial conditions for $x$ and $y$ at $t=0$ (they are mutually mirror imaged \cite{takahashi_quantization_2018}).

We can obtain the initial conditions for the quantum variables $G^{a,b,c,d}_{1}$ by applying an inverse transformation, and rewriting them in terms of quantum variables for which we can use the coherent states. For instance, for $G_{1}^{2,0,0,0}$ we get
\begin{equation}
 G^{2,0,0,0}_{1} = \langle(\hat{x}_{1}-\langle\hat{x_{1}}\rangle)^{2} \rangle_{\text{Weyl}} =\frac{1}{2} \langle \left[ (\hat{x}-\langle\hat{x}\rangle)+(\hat{y}-\langle\hat{y}\rangle) \right]^{2} \rangle_{\text{Weyl}}
    =\frac{1}{2}\Big[G^{2,0,0,0}+G^{0,0,2,0}+2G^{1,0,1,0}\Big].    
\end{equation}

Given that in the classical description there is an extra degree of freedom representing the mirror image of $x$ and $p_{x}$, we assume a similar situation in the quantum counterpart. This means equal initial conditions for the quantum variables $(y, p_{y})$ and $(x, p_{x})$. Explicitly, from Eq. 
 (\ref{eq:initial_con_quantum_var})
\begin{align}
    &G^{2,0,0,0}_{t=0}=\langle \hat{x}^{2} \rangle_{t=0} -\langle \hat{x} \rangle_{t=0}^{2}=\frac{\hbar}{2m\omega},\nonumber\\
    &G^{0,2,0,0}_{t=0}=\langle \hat{p}_{x}^{2} \rangle_{t=0} -\langle \hat{p}_{x} \rangle_{t=0}^{2}=\frac{m\hbar\omega}{2},\nonumber\\ 
    &G^{1,1,0,0}_{t=0}=\big(\langle \hat{x}\hat{p}_{x} \rangle_{\text{Weyl}}\big)_{t=0} -\langle \hat{x} \rangle \langle \hat{p}_{x} \rangle_{t=0}=0,\nonumber \\
    &G^{0,0,2,0}_{t=0}=\langle \hat{y}^{2} \rangle_{t=0} -\langle \hat{y} \rangle_{t=0}^{2}=\frac{\hbar}{2m\omega},\nonumber\\
    &G^{0,0,0,2}_{t=0}=\langle \hat{p}_{y}^{2} \rangle_{t=0} -\langle \hat{p}_{y} \rangle_{t=0}^{2}=\frac{m\hbar\omega}{2},\nonumber\\ 
    &G^{0,0,1,1}_{t=0}=\big(\langle \hat{y}\hat{p}_{y} \rangle_{\text{Weyl}}\big)_{t=0} -\langle \hat{y} \rangle \langle \hat{p}_{y} \rangle_{t=0}=0,
\end{align}
and by using the inverse transform relations, we get the initial conditions for the SBTH
\begin{align} \label{initial_Gs}
        &x_{10}=2\sqrt{\frac{n\hbar}{m\omega}} & &x_{20}=0 & &p_{10}=0 & &p_{20}=0 \nonumber \\
        &G^{2,0,0,0}_{1_{t=0}}=\frac{\hbar}{2m\omega} & &G^{0,2,0,0}_{1_{t=0}}=\frac{m\hbar\omega}{2} &
        &G^{0,0,2,0}_{1_{t=0}}=\frac{m\hbar\omega}{2} &
        &G^{0,0,0,2}_{1_{t=0}}=\frac{\hbar}{2m\omega} \nonumber \\
        &G^{1,1,0,0}_{1_{t=0}}=0 & &G^{1,0,1,0}_{1_{t=0}}=0 &
        &G^{1,0,0,1}_{1_{t=0}}=0 & &G^{0,1,0,1}_{1_{t=0}}=0 \nonumber \\
        &G^{0,0,1,1}_{1_{t=0}}=0 & &G^{0,1,1,0}_{1_{t=0}}=0
\end{align}

\subsection{Dynamical Evolution}

As we mentioned in section \ref{sec:semiclassical_equation_BT},
the dynamics for our SBTH model, Eq. (\ref{eq:G_QDHO}), and that of  Lindblad's model, Eq. (\ref{eq:G_Lindblad}), are completely equivalent at the effective level; classical and quantum variables have similar equations of motion, Eqs. (\ref{eq:G_QDHO}-\ref{eq:quantum_equality}). However, it is important to remember that although the equations of motion are classical (obtained from a Hamiltonian), the whole system is quantum in nature, so its probabilistic behavior must be tracked. We show next how to do this.

We use the initial conditions proposed in the previous section to obtain the dynamical evolution of the SBTH and Lindblad's SDEs, and the following parameters
\begin{equation}
    x_{0}=2\sqrt{\frac{n\hbar}{m\omega}}, \quad p_{x_{0}}=0\quad \gamma=0.08, \quad \omega=\omega_{0}'=1.5, \quad \bar{n}\in\{0,1,2\} \quad m=\hbar=1, \quad n=3.
\end{equation}

\begin{figure}[h]
    \centering
    \includegraphics[width=15cm]{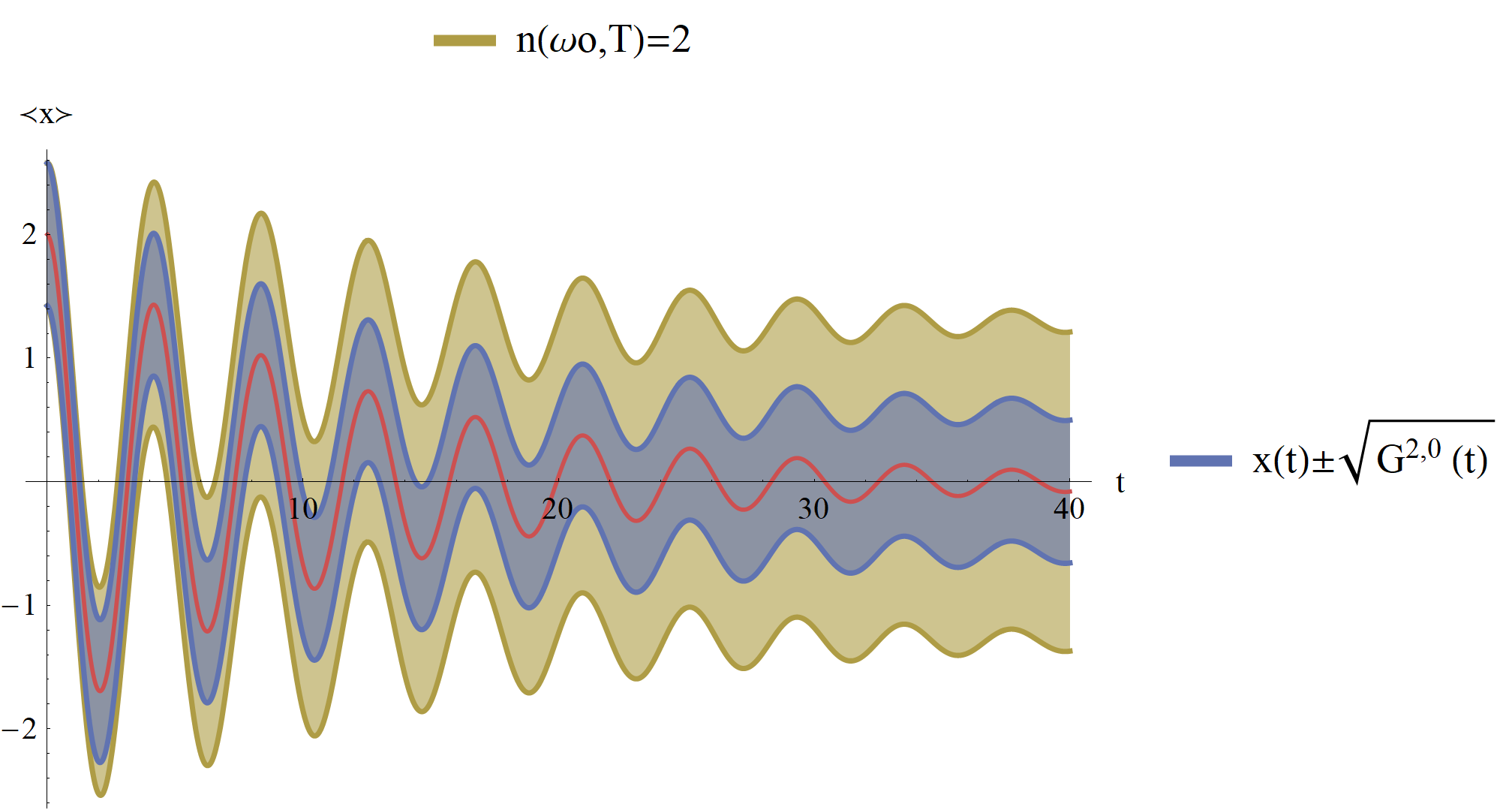}
    \caption{Evolution of $\langle\hat{x}(t)\rangle$ of the QDHO (red line), and its dispersion belts: Lindblad's with $\bar{n}=2$ (golden shaded area), and SBTH (purple shaded area).}
    \label{fig:position_dispersion}
\end{figure}
\begin{figure}[h]
    \centering
    \includegraphics[width=15cm]{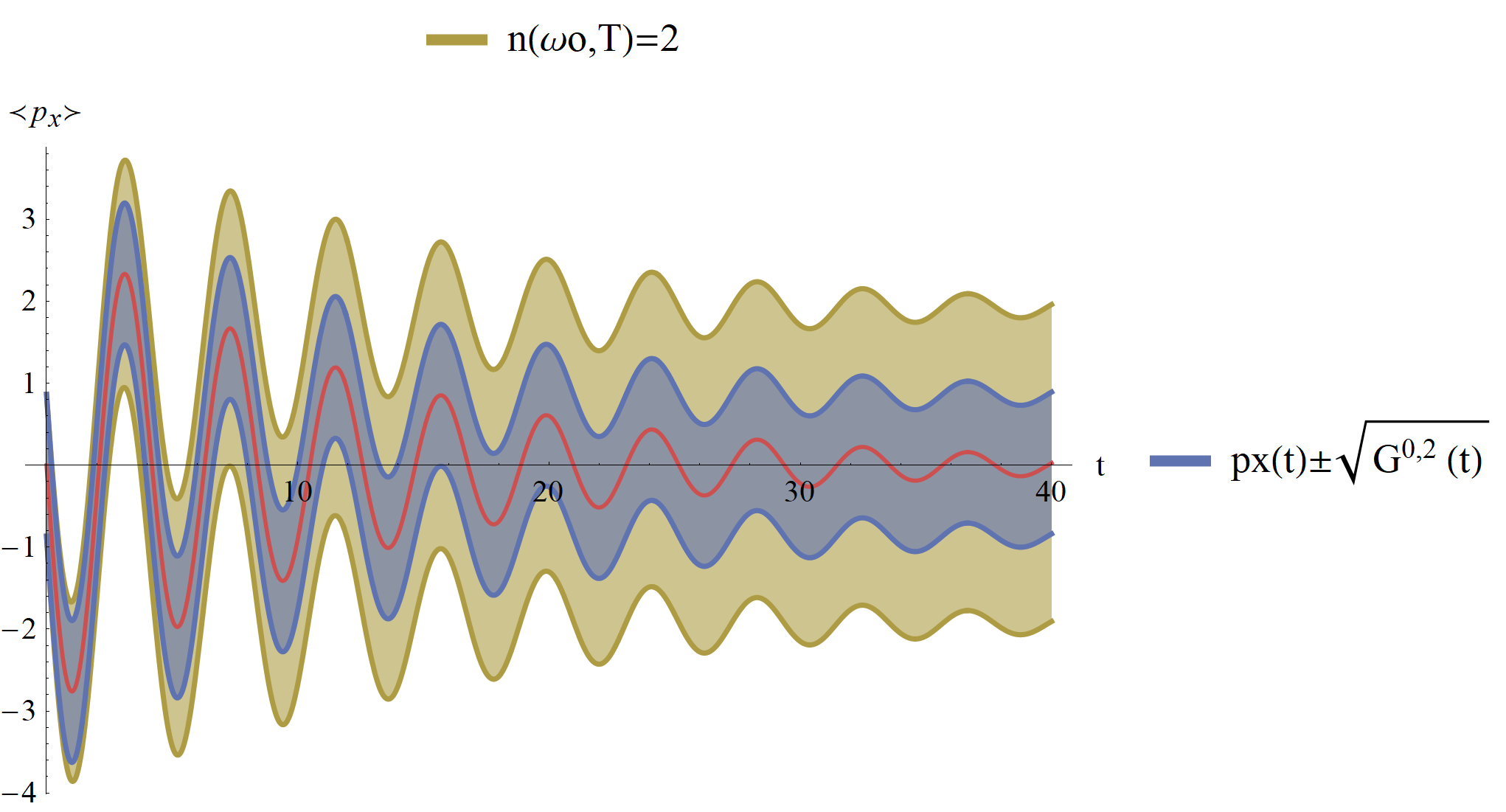}
    \caption{Evolution of $\langle\hat{p}_x (t)\rangle$ of the QDHO (red line), and its dispersion belts: Lindblad's with $\bar{n}=2$ (golden shaded area), and SBTH (purple shaded area).}
    \label{fig:momentum_dispersion}
\end{figure}
\begin{figure}[h]
    \includegraphics[width=17cm]{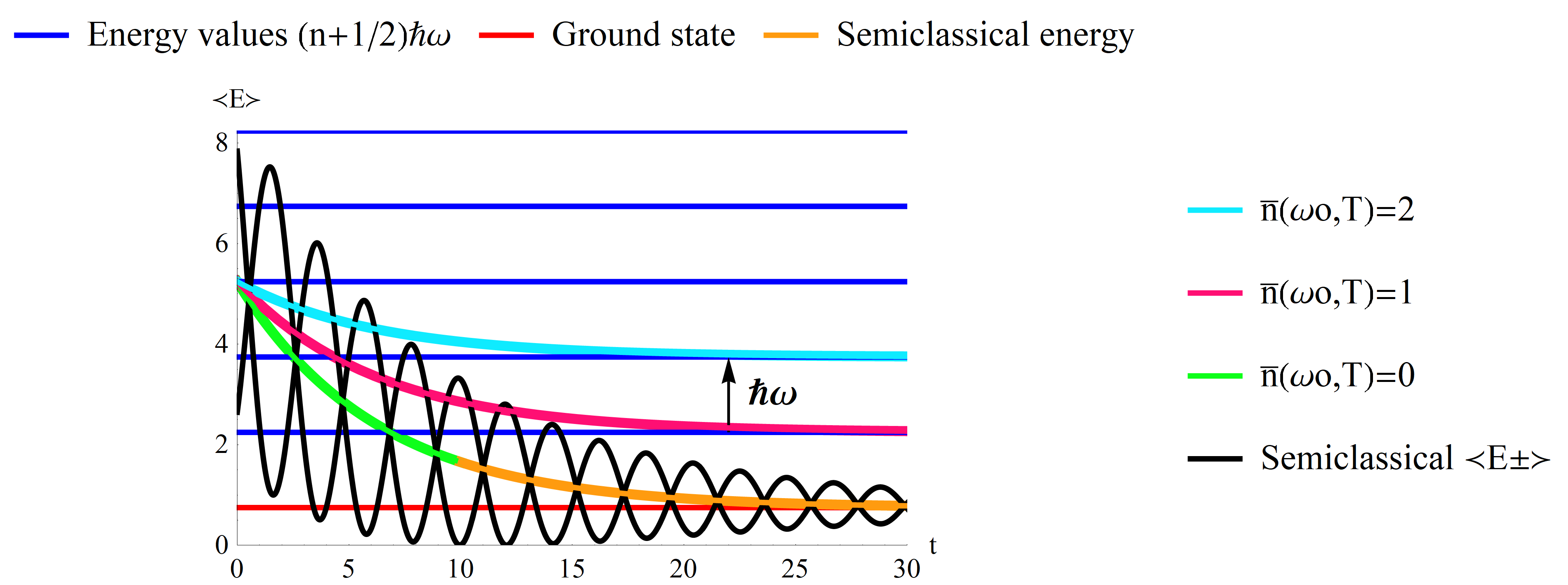}
    \caption{Energy evolution of the QDHO for $\bar{n}=0,1,2$ (brown, pink, and blue solid lines), and the QDHO's semiclassical energy (orange dashed line), together with its dispersion energy belt (black solid lines).}
    \label{fig:QDHO's energy}
\end{figure}

 As we mentioned above, the behavior of physical variables should display their probabilistic nature, hence, for a physically meaningful evolution, quantum dispersions should be taken into account, that is, \[ x(t) \pm \Delta x = x(t) \pm \sqrt{G^{2,0}(t)},\quad
 p(t)\pm\Delta p=p(t)\pm \sqrt{G^{0,2}(t)}.\] 
 
 In Figs. \ref{fig:position_dispersion} and \ref{fig:momentum_dispersion}, we show the evolution of position and momentum, alongside their uncertainty regions bounded by their corresponding dispersions, for both SBTH (purple shaded area) and Lindblad (golden shaded area), respectively. 
 Evidently, the dispersions exhibit non-trivial dynamics within the effective approach, yet their evolution remains distinctly synchronized with the classical variables. Furthermore, one can observe that the amplitudes of the dispersions differ between the two descriptions due to the thermal values  $\bar{n}$ in Lindblad's approach. In contrast, in our description, the uncertainty regions remain consistently distanced from the semiclassical evolution.

Although both approaches yield similar results, our semiclassical method has a clear advantage over other schemes: it is possible to determine the quantum mechanical behavior of any physical variable in a direct, more intuitive way. In particular, we can analyze the energy of the QDHO: the classical total mechanical energy of the harmonic oscillator is given by
\begin{align}
    E_{mech}&=K(t)+U(t)\nonumber\\
    &=\frac{\big(p_{x}(t)\big)^{2}}{2m}+\frac{1}{2}m\omega^{2}\big(x(t)\big)^{2}.
\end{align}
This mechanical energy corresponds to the Hamiltonian, and thus we extend the  expression above to the quantum energy within the momentous formalism
\begin{equation}
\label{eq:mec_energ}
    \langle E(t)\rangle=\frac{1}{2m}\big(p_{x}(t)\big)^{2}+\frac{1}{2}m\omega^{2}\big(x(t)\big)^{2}+\frac{1}{2m}G^{0,2}(t)+\frac{1}{2}m\omega^{2}G^{2,0}(t).
\end{equation}
We can also give an expression including the quantum uncertainty in the evolution as follows
\begin{equation}
\label{eq:Energy_evolution_momentous}
    \langle E_{\pm} \rangle=\frac{1}{2m}\Big(p_{x}(t)\pm \sqrt{G^{0,2}(t)}\Big)^{2}+\frac{1}{2}m\omega^{2}\Big(x(t)\pm \sqrt{G^{2,0}(t)}\Big)^{2}.
\end{equation}

Eq. (\ref{eq:mec_energ}) represents the mean energy of the system within the momentous quantum mechanics scheme, being an effective alternative to Eq. (\ref{eq:mean_energy_lindblad}) in the Lindblad approach.

In Fig. (\ref{fig:QDHO's energy}), we obtained QDHO's mean energy as described in Eq. (\ref{eq:mec_energ}). Note how the semiclassical trajectory (orange dashed line) and the one in the Lindblad approach (brown solid line), Eq. (\ref{eq:mean_energy_lindblad}), coincide for $\bar{n}=0$. 
Furthermore, one can see the decay of the QDHO's energy: initially, the QDHO's energy is in the third excited state $n=3$, and as it evolves in time, this energy gradually converges towards the ground state.

\section{Discussion and conclusions}
In this study, we have derived an effective description of the quantum damped harmonic oscillator, wherein semiclassical Hamiltonian equations govern the evolution of expectation values for physical operators. We have shown how the challenging issue of energy dissipation in quantum mechanics can be implemented and analyzed, particularly for the harmonic oscillator. We hope our contributions will pave the way for research into complex phenomena and general open quantum systems.
As we mentioned in the text, we contrasted our results with those obtained in the Lindblad formulation with master equations, demonstrating the robustness of our approach.

The effective evolution shown in the preceding section precisely illustrates this point, revealing remarkable similarities between the dynamics derived from the SBTH and the Lindblad approach in the study of the QDHO.
From this, we understand that the SBTH description not only overcomes the shortcomings and difficulties encountered in the canonical quantization of the BTH, but it also exhibits the same behavior of the QDHO obtained through master equations. The solution given by employing a coherent state shows that the initial quantum state for the semiclassical model is preserved throughout its evolution. Explicitly, the effective dynamics of the QDHO is governed by Eq. (\ref{eq:G_QDHO}), that, fed with initial conditions Eq. (\ref{initial_Gs}), provides the evolution of the system, in perfect agreement with Lindblad's Eq. (\ref{eq:G_Lindblad}) for $\bar{n}=0$.

To the authors' knowledge, no prior explicit comparison between the Bateman model and the Lindblad master equation for the QDHO has been obtained. Remarkably, the success of our effective description was achieved mainly due to the introduction of the canonical transformations Eqs. (\ref{eq:class_bateman2})-(\ref{bateman2}), and the independent nature of classical and quantum variables Eq. (\ref{eq:difference_quantum_classical}).

Naive straightforward canonical quantization of the Bateman model results in an invalid evolution, which is corrected by using the canonical transformation mentioned above, which preserves the Heisenberg uncertainty and describes the same quantum dynamics as the one provided by Lindblad's, as demonstrated by the quantization procedure shown in section \ref{sec:semiclassical_equation_BT}.

Furthermore, we can demonstrate that the presence of a ground state arises as a consequence of the generalized uncertainty constraint applied to the quantum variables (\ref{eq:uncertainty}). Analyzing the evolution of the quantum effective energy, Eq. (\ref{eq:mec_energ}), the contribution of $x(t)$ and $p(t)$ for large times is negligible, as classical variables should do. Thus we end up with
\begin{equation}
    \langle E\rangle=\frac{1}{2m}G^{0,2}+\frac{1}{2}m\omega^{2}G^{2,0}.
\end{equation}
We can find a more suitable expression by multiplying by $G^{2,0}$ on both sides of this equation
\begin{equation}
    G^{2,0}\langle E\rangle=\frac{1}{2m}G^{2,0}G^{0,2}+\frac{1}{2}m\omega^{2}(G^{2,0})^{2}.
\end{equation}
Now, by using the uncertainty relation $G^{2,0}G^{0,2}\geq\hbar^{2}/4$, we get
\begin{equation}
   G^{2,0}\langle E\rangle\geq\frac{1}{2m}\frac{\hbar^{2}}{4}+\frac{1}{2}m\omega^{2}(G^{2,0})^{2}, 
\end{equation}
or, in a more insightful way
\begin{equation}
   \frac{1}{2}m\omega^{2}(G^{2,0})^{2}-G^{2,0}\langle E\rangle+\frac{1}{2m}\frac{\hbar^{2}}{4} \leq 0. 
\end{equation}
Here we observe that $G^{2,0}$ is equal or greater than the minimum value given by the solution of the quadratic equation
\begin{equation}
    G^{2,0}_{\pm}=\frac{1}{m\omega^{2}}\Big(\langle E\rangle\pm \sqrt{\langle E\rangle^{2}-\hbar^{2}\omega^{2}/4}\Big),
\end{equation}
however, by definition, quantum variables satisfy the following conditions
\begin{equation}
    \text{Im}[G^{2,0}]=0\quad \text{and}\quad G^{2,0}>0.
\end{equation}
Finally we get
\begin{equation}
    \langle E\rangle^{2}\geq\frac{\hbar^{2}\omega^{2}}{4}\;\Rightarrow\;\langle E\rangle\geq \frac{\hbar\omega}{2},
\end{equation}
thereby confirming the existence of a ground state.

We conclude by emphasizing that the momentous effective approach captures the essential information of quantum systems,  which, as demonstrated in the present manuscript, can accommodate dissipative effects. It is also important to mention that the particular expression of the effective Hamiltonian (\ref{eq:quantum_corrected}) for the damped oscillator is, in general, much more involved, usually containing an infinite number of quantum corrections \cite{doi:10.1063/1.4748550,Quantum_Cosmology}; the evolution in such cases can be analyzed order by order in the quantum variables, or by the implementation of an alternative description in terms of canonical variables \cite{PhysRevA.99.042114}. Extension to more general open quantum systems will appear in future work.

\begin{flushleft}
		\begin{Large}
			\textbf{\hypertarget{appendix2}{Appendix}: Poisson Brackets between Quantum Dynamical Variables}
		\end{Large}
	\end{flushleft}

The Poisson bracket between quantum variables, for two degrees of freedom, $\{G^{a,b,c,d},G^{e,f,g,h}\}$\footnote{We are considering the moments defined in Eq. (\ref{Generalized effective dynamical variables})}, can be computed by using  \cite{PhysRevD.31.1341,ashtekar_geometrical_1997}
\begin{equation}
\label{eq:Heslot_bracket}
    \{\langle\hat{f}\rangle,\langle\hat{g}\rangle\}=\frac{1}{i\hbar}\langle[\hat{f},\hat{g}]\rangle.
\end{equation}

In the momentous approach, as mentioned in section \ref{section:formalism_momentous}, equations of motion can be obtained in the usual way, $\dot{q}_{i}=\{q_{i},H_Q\}$. For the SBTH, the effctive Hamiltonian reads
\begin{align}
     H_{Q} =&\left(\frac{p_{1}^{2}}{2m}+\frac{1}{2}m\Omega^{2}x_{1}^{2}\right)-\left(\frac{p_{2}^{2}}{2m}+\frac{1}{2}m\Omega^{2}x_{2}^{2}\right)-\lambda(x_{1}p_{2}+x_{2}p_{1})-\lambda G^{1,0,1,0}_{1}\nonumber\\
    &+\frac{m\Omega^{2}}{2}G^{2,0,0,0}_{1}+\frac{1}{2m}G^{0,2,0,0}_{1}-\frac{m\Omega^{2}}{2}G^{0,0,0,2}_{1}-\frac{1}{2m}G^{0,0,2,0}_{1}-\lambda G^{0,1,0,1}_{1}
\end{align}
the equations of motion are given by
\begin{equation}
\begin{split}
    \dot{x}_{k}=\{x_{k},H_{Q}\}\quad\dot{p}_{k}=\{p_{k},H_{Q}\}
\end{split}
\end{equation}
and
\begin{equation}
\label{eq:moments_equation}
    \dot{G}_{1}^{a,b,c,d}=\{G_{1}^{a,b,c,d},H_{Q}\}
\end{equation}
where $k=\{1,2\}$. 


The non-trivial Poisson brackets between momenta $\{G^{a,b,c,d}_{1},G^{e,f,g,h}_{1}\}$ are the following %
\begin{align*}
&\{G_{1}^{2,0,0,0},G_{1}^{1,0,1,0}\}=0   &   &\{G_{1}^{1,0,0,1},G_{1}^{0,0,2,0}\}=-2G_{1}^{1,0,1,0}\\
&\{G_{1}^{2,0,0,0},G_{1}^{0,1,0,1}\}=2G_{1}^{1,0,0,1} & &\{G_{1}^{0,1,1,0},G_{1}^{1,0,1,0}\}=-G_{1}^{0,0,2,0}\\
&\{G_{1}^{2,0,0,0},G_{1}^{0,2,0,0}\}=4G_{1}^{1,1,0,0} & &\{G_{1}^{0,0,1,1},G_{1}^{0,0,0,2}\}=2G_{1}^{0,0,0,2}\\
&\{G_{1}^{0,2,0,0},G_{1}^{1,0,1,0}\}=-2G_{1}^{0,1,1,0} & &\{G_{1}^{0,1,1,0},G_{1}^{0,1,0,1}\}=G_{1}^{0,2,0,0}\\
&\{G_{1}^{0,0,2,0},G_{1}^{0,1,0,1}\}=2G_{1}^{0,1,1,0} & &\{G_{1}^{0,1,1,0},G_{1}^{2,0,0,0}\}=-2G_{1}^{1,0,1,0}\\
&\{G_{1}^{0,0,2,0},G_{1}^{0,0,0,2}\}=4G_{1}^{0,0,1,1} & &\{G_{1}^{0,1,1,0},G_{1}^{0,0,0,2}\}=2G_{1}^{0,1,0,1}\\
&\{G_{1}^{0,0,0,2},G_{1}^{1,0,1,0}\}=-2G_{1}^{1,0,0,1} & &\{G_{1}^{1,1,0,0},G_{1}^{1,0,1,0}\}=-G_{1}^{1,0,1,0}\\
&\{G_{1}^{1,1,0,0},G_{1}^{0,1,0,1}\}=G_{1}^{0,1,0,1} & &\{G_{1}^{1,1,0,0},G_{1}^{0,2,0,0}\}=2G_{1}^{0,2,0,0}\\
&\{G_{1}^{0,1,0,1},G_{1}^{2,0,0,0}\}=-2G_{1}^{1,0,0,1} & &\{G_{1}^{1,1,0,0},G_{1}^{2,0,0,0}\}=-2G_{1}^{2,0,0,0}\\
&\{G_{1}^{1,0,0,1},G_{1}^{1,0,1,0}\}=-G_{1}^{2,0,0,0} & &\{G_{1}^{0,0,1,1},G_{1}^{1,0,1,0}\}=-G_{1}^{1,0,1,0}\\
&\{G_{1}^{1,0,0,1},G_{1}^{0,1,0,1}\}=G_{1}^{0,0,0,2} & &\{G_{1}^{0,0,1,1},G_{1}^{0,1,0,1}\}=G_{1}^{0,1,0,1}\\
&\{G_{1}^{1,0,0,1},G_{1}^{0,2,0,0}\}=2G_{1}^{0,1,0,1} & &\{G_{1}^{0,0,1,1},G_{1}^{0,0,2,0}\}=-2G_{1}^{0,0,2,0}\\
&\{G_{1}^{1,0,0,1},G_{1}^{0,1,1,0}\}=G_{1}^{0,0,1,1}-G_{1}^{1,1,0,0} & &\{G_{1}^{1,0,1,0},G_{1}^{0,1,0,1}\}=G_{1}^{1,1,0,0}+G_{1}^{0,0,1,1}.
\end{align*}
%
%
%

The explicit calculation is, for example, as follows
\begin{align}
    \{G_{1}^{2,0,0,0},G_{1}^{0,2,0,0}\}&=(i\hbar)^{-1}\langle[(\hat{x}_{1}-x_{1})^{2},(\hat{p}_{1}-p_{1})^{2}]\rangle\nonumber\\
    &=(i\hbar)^{-1}\langle(\hat{x}_{1}-x_{1})[\hat{x}_{1},\hat{p}_{1}](\hat{p}_{1}-p_{1})+[\hat{x}_{1},\hat{p}_{1}](\hat{x}_{1}-x_{1})(\hat{p}_{1}-p_{1})\nonumber\\
    &+(\hat{p}_{1}-p_{1})(\hat{x}_{1}-x_{1})[\hat{x}_{1},\hat{p}_{1}]+(\hat{p}_{1}-p_{1})[\hat{x}_{1},\hat{p}_{1}](\hat{x}_{1}-x_{1})\rangle\nonumber\\
    &=2\langle(\hat{x}_{1}-x_{1})(\hat{p}_{1}-p_{1})+(\hat{p}_{1}-p_{1})(\hat{x}_{1}-x_{1})\rangle\nonumber\\
    &=4\langle(\hat{x}_{1}-x_{1})(\hat{p}_{1}-p_{1})\rangle_{\text{Weyl}}\nonumber\\
    &=4G_{1}^{1,1,0,0}.
\end{align}
and similarly for the rest.

\printbibliography

\end{document}